\documentclass[a4paper,12pt]{article}

\usepackage{amstext,amsmath,amssymb,amsthm}  % AMS Math
\usepackage[normalem]{ulem}  % For crossing out text using \sout{}
\usepackage{graphicx}  % For figures
\usepackage{booktabs}  % For beautiful tables
\usepackage{bm} % For bold greek letters and mathematical symbols

\usepackage[comma,sort,authoryear]{natbib}
\usepackage{hypernat}  % For natbib to interact with hyperref

\usepackage{vmargin}  % Defaults margins to sensible values on A4 paper
\usepackage{authblk}

\usepackage{xcolor}
\definecolor{dark-red}{rgb}{0.8,0.15,0.15}
\definecolor{dark-blue}{rgb}{0.15,0.15,0.6}
\definecolor{medium-blue}{rgb}{0,0,0.8}

\usepackage[pdftex,breaklinks,colorlinks]{hyperref}
\hypersetup{  % hyperref configuration, metadata
  pdftitle =
    {The average bond-length of diatomic molecules in thermodynamical equilibrium depends on the volume},
  linkcolor={dark-red},
  citecolor={dark-blue},
  urlcolor={medium-blue}
}

\usepackage{pgfplots}
\pgfplotsset{compat=1.18, width=8cm}

\begin{document}

\numberwithin{equation}{section}
\numberwithin{figure}{section}
\allowdisplaybreaks[1]  % To allow eqnarrays to break between pages

\title{The average bond-length of diatomic molecules in thermodynamical equilibrium depends on the volume}

\author[1]{Pablo Echenique-Robba}

\affil[1]{Instituto de Qu\'{\i}mica F\'{\i}sica Blas Cabrera, CSIC, Madrid, Spain}

\date{\today}

\maketitle

% With an abstract
\begin{abstract}

In the framework of classical statistical mechanics and assuming the electronic ground-state Born-Oppenheimer approximation, we show in this work that a dependence of the equilibrium bond-length on the available volume is to be expected for a dilute gas of diatomic molecules. In a nutshell, this dependence is controlled by the relation between the potential well depth $D$ (through the factor $e^{-\beta D}$) and the quotient $L/R$ of the linear size of the container $L$ and the potential well width $R$. Using simplified analytical estimations, we predict that the equilibrium bond-length $\langle \rho \rangle$ is independent of $L$ for a range of volumes which is exponentially large on $\beta D$. At some point of the $L$ axis, $\langle \rho \rangle$ starts to increase and it eventually diverges, as it is to be expected, when $L \to \infty$, thus describing dissociated atoms. According to our estimations, it is possible that some diatomic molecules, such as halogens, could present volume dependence of their equilibrium bond-lengths for laboratory-size volumes at room temperature 
\vspace{0.4cm}
\vspace{0.2cm}\\

\end{abstract}

\section{Introduction to the problem}
\label{sec:intro}

Say we have a dilute gas of $N \sim N_\mathrm{A}$ diatomic molecules, being $N_\mathrm{A}$ Avogadro's number. Say we denote by $m_a$ and $m_b$ the masses of atoms of types $a$ and $b$, respectively, by $\rho_i$ the bond-length of the $i$-th molecule, with $i=1,\ldots,N$, and by $x_{a,i}^k$ and $x_{b,i}^k$, with $k=1,2,3$, the Euclidean coordinates of atoms of types $a$ and $b$, respectively, of the $i$-th molecule.

Then the average equilibrium bond-length in the microcanonical (i.e., constant $T$) ensemble is:
\begin{eqnarray}
\label{eq:r_av_Euc}
\left\langle \frac{1}{N} \sum_i \rho_i \right\rangle & = &
 \frac{\displaystyle \int \left( \frac{1}{N} \sum_i \rho_i \right)
    \exp{\left[-\beta \sum_i V(\rho_i)\right]} 
	\,\left( \prod_i \prod_{k=1}^3 \mathrm{d}x_{a,i}^k
	                               \mathrm{d}x_{b,i}^k \right)}
    {\displaystyle \int \exp{\left[-\beta \sum_i V(\rho_i)\right]} 
	\,\left( \prod_i \prod_{k=1}^3 \mathrm{d}x_{a,i}^k
	                               \mathrm{d}x_{b,i}^k \right)}
    \nonumber \\
 & = & \frac{\displaystyle \int \rho \exp{\left[-\beta V(\rho)\right]} 
	\,\left( \prod_{k=1}^3 \mathrm{d}x_a^k
	                       \mathrm{d}x_b^k \right)}
    {\displaystyle \int \exp{\left[-\beta V(\rho)\right]} 
	\,\left( \prod_{k=1}^3 \mathrm{d}x_a^k
	                       \mathrm{d}x_b^k \right)}
   =: \langle \rho \rangle \ ,
\end{eqnarray}
where the following assumptions have been made:

\begin{itemize}
\item {\bf Negligible interaction between the different molecules} (i.e., the dilute gas approximation). This is which allows us to write the potential energy term as a sum of the individual molecular potential energies.
\item {\bf The Born-Oppenheimer approximation} (i.e., the approximation at the level of quantum mechanics that electrons experience the presence of nuclei as if they were point charges clamped at each position and nuclei experience the presence of electrons through the potential energy surface associated to the average electronic energy at each electronic quantum level). For more technical information about this assumption, see \cite{Echenique2007a}.
\item {\bf The ground-state approximation for the electrons}, as it is usual in the quantum chemistry field. Thus, the potential energy surface produced by the ground-state electrons on the nuclei is $V(\rho)$. Since we also assume that there is no external field applied on the system, this energy can only depend on the bond-length $\rho$.
\item {\bf The classical mechanics approximation for the nuclei.} This is why we write the classical statistical mechanics average in eq.~(\ref{eq:r_av_Euc}) instead of its more general quantum counterpart. This approximation will be more accurate when the nuclei are heavier and the temperature higher. For lighter nuclei and lower temperatures, quantum corrections should be explored in further works.
\end{itemize}

Under these assumptions, the result in eq.~(\ref{eq:r_av_Euc}) is exact. In order to arrive to it, we have taken into account the cancellation of a number of coordinate-independent quantities appearing both at the numerator and the denominator, including the ones produced by the integration of the Euclidean momenta, which can be performed analytically simply using the well-known Gaussian integration formula.

In order to arrive to even simpler expressions, we can change coordinates to a special flavor of sphericals that is often used in the simulation of biological molecules. Let us denote by $(X,Y,Z)$ the Euclidean coordinates of atom $a$, by $\rho$ the corresponding bond-length, and by $\theta$ and $\phi$ the two angles that are needed to completely specify the position of atom $b$. Then the change of coordinates is given by:
\begin{subequations}
\label{eq:change_of_coords}
\begin{align}
& x^1_a = X \  , \label{eq:xa1} \\
& x^2_a = Y \  , \label{eq:xa2} \\
& x^3_a = Z \  , \label{eq:xa3} \\
& x^1_b = X + \rho \sin \theta \cos \phi \  , \label{eq:xb1} \\
& x^2_b = Y + \rho \sin \theta \sin \phi \  , \label{eq:xb2} \\
& x^3_b = Z + \rho \cos \theta \  . \label{eq:xb3}
\end{align}
\end{subequations}

Now, we can change coordinates under the integral sign in eq.~(\ref{eq:r_av_Euc}) through the substitution of the volume element as follows:
\begin{equation}
\label{eq:change_of_coords_int}
\left( \prod_{k=1}^3 \mathrm{d}x_a^k \mathrm{d}x_b^k \right)
 \ \longrightarrow \
\det J(\bm{q}) \left( \prod_{j=1}^6 \mathrm{d}q^j \right) \ ,
\end{equation}
where $\bm{q}:=(X,Y,Z,\rho,\theta,\phi)$ is the full set of new, curvilinear coordinates, and $\det J(\bm{q})$ is the Jacobian determinant, which, in this particular case, can be explicitly calculated:
\begin{eqnarray}
\label{eq:detJ}
\det J(\bm{q}) & := & \det \left( \begin{array}{cccccc}
       1 & 0 & 0 & 1 & 0 & 0 \\
       0 & 1 & 0 & 0 & 1 & 0 \\
       0 & 0 & 1 & 0 & 0 & 1 \\
       0 & 0 & 0 & \sin\theta \cos\phi & \sin\theta \sin\phi & \cos\theta \\
       0 & 0 & 0 & \rho \cos\theta \cos\phi & \rho \cos\theta \sin\phi & -\rho \sin\theta \\
       0 & 0 & 0 & -\rho \sin\theta \sin\phi & \rho \sin\theta \cos\phi & 0
\end{array} \right) \nonumber \\[5pt]
               & = & \rho^2 \sin^3\theta \sin^2\phi
                   + \rho^2 \cos^2\theta \sin\theta \cos^2\phi 
				   \nonumber \\
			   & & + \rho^2 \cos^2\theta \sin\theta \sin^2\phi
				   + \rho^2 \sin^3\theta \cos^2\phi
				   \nonumber \\
			   & = & \rho^2 \sin^3\theta
			       + \rho^2 \cos^2\theta \sin\theta
				   \nonumber \\
			   & = & \rho^2 \sin\theta \ .
\end{eqnarray}

The angles $\theta$ and $\phi$ can be analytically integrated out in eq.~(\ref{eq:r_av_Euc}), yielding identical constant factors at the numerator and the denominator it, and finally producing a 1-dimensional integral over the bond-length $\rho$:
\begin{equation}
\label{eq:r_av_sph}
\left\langle \frac{1}{N} \sum_i \rho_i \right\rangle
 = \langle \rho \rangle 
 = \frac{\displaystyle \int \rho^3 \exp{\left[-\beta V(\rho)\right]} 
	\,\mathrm{d}\rho}
    {\displaystyle \int \rho^2 \exp{\left[-\beta V(\rho)\right]} 
	\,\mathrm{d}\rho} \ .
\end{equation}

Now, it is easy to see that, for potentials $V(\rho)$ that satisfy
\begin{subequations}
\label{eq:Vfamily}
\begin{align}
\lim_{\rho \to 0} V(\rho) & = V^0 \ , \label{eq:Vfamily_a} \\
\lim_{\rho \to \infty} V(\rho) & = V^\infty \ , \label{eq:Vfamily_b}
\end{align}
\end{subequations}
where $V^0$ can be a number or $\infty$, and $V^\infty$ is a number, i.e., for potentials that are well-behaved in $\rho=0$ and have a horizontal asymptote in $\rho=\infty$, not only are both improper integrals in eq.~(\ref{eq:r_av_sph}) divergent if we assume that the integration limits for $\rho$ are $(0,\infty)$, but their quotient, $\langle \rho \rangle$, is divergent too.

This is the starting point of our analysis.

\section{Morse potential and interstellar Hydrogen}
\label{sec:H}

First of all, we have to stress that, although it may seem at first sight that the divergence described above could be problematic or counterintuitive, it is in fact a qualitatively correct result.

\begin{figure}[!ht]
\begin{center}
\begin{tikzpicture}
\begin{axis}[xmin=0, xmax=10, ymin=0, ymax=10,
	clip=false,
	axis lines=left, axis line style={very thick},
	xtick={2}, xticklabels={$\rho_\mathrm{eq}$},
	ytick={5}, yticklabels={$V^\infty$},
	xmajorgrids=true, ymajorgrids=true, grid style=dashed,
	x label style={at={(axis description cs:0.95,-0.01)},anchor=north},
	y label style={at={(axis description cs:-0.1,0.85)},rotate=-90,anchor=south},
	xlabel=$\rho$, ylabel=$V(\rho)$]
\addplot[color=red, very thick, domain=1.13:9, samples=50]{5*(1-(e^(2-x)))^2};
\end{axis}
\end{tikzpicture}
\caption{\label{fig:Morse}{\small Schematic depiction of the Morse model potential given by eq.~(\ref{eq:V_Morse}).}}
\end{center}
\end{figure}
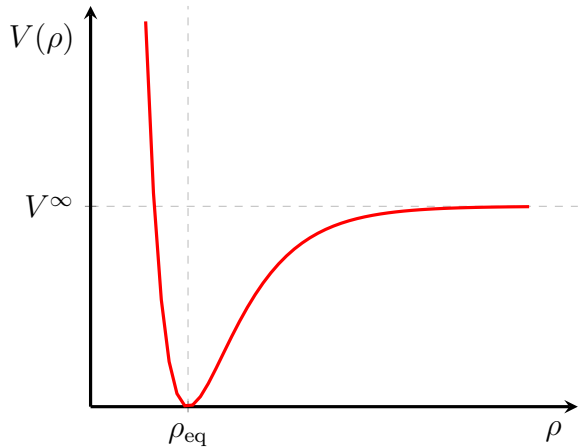

On the one hand, the requirements in eq.~(\ref{eq:Vfamily}) for the potential energy as a function of the interatomic distance $r$ are not only very reasonable (large or infinite energy values when the nuclei get too close and the Coulombic repulsion between them becomes divergent, and no force ---i.e., constant energy--- when they are very far apart), but they are also present in simple, analytical, model potentials which are well-known to mimic the exact calculation of the electronic plus Coulombic potential energy function in the Born-Oppenheimer approximation such as the Morse potential \citep{Morse1929} depicted in fig.~(\ref{fig:Morse}) and given by:
\begin{equation}
\label{eq:V_Morse}
V_\mathrm{Morse}(\rho) := V^\infty \left( 
  1 - e^{-a(\rho-\rho_\mathrm{eq})}
  \right)^2 \ ,
\end{equation}
where
\begin{equation}
\label{eq:a}
a = \sqrt{\frac{k}{2V^\infty}} ,
\end{equation}
being $k$ the force constant at the equilibrium $\rho=\rho_\mathrm{eq}$.

As one can easily see, this model potential satisfies the requirements in eq.~(\ref{eq:Vfamily}). Its value for $\rho=0$ is not infinite, but it can be shown to be very large for typical diatomic molecules.

If we now assume the basic principle of ergodicity in statistical thermodynamics, it becomes then obvious that, in eq.~(\ref{eq:r_av_sph}), the integration region corresponding to very large bond-length is infinitely larger than the region for which $\rho$ is of molecular magnitude, thus overcoming the average and yielding a thermodynamical bond-length which describes a dissociated molecule with the nuclei divergently far apart from each other.

Although proper calculations should include quantum-mechanical effects, this qualitative result is consistent too with the fact that most of interstellar Hydrogen exists in its neutral, atomic form, as opposed to the H$_2$ molecular state (which also exists, but mostly within colder, denser interstellar clouds) \citep{Verschuur1988}.

This being said, while it is reasonable and intuitive to expect that diatomic molecules are dissociated into their atomic form in the vast regions of almost empty outer space, our everyday laboratory experience with diatomic gases is not at all that one. Normally, diatomic molecules that are at hand are stable entities with more or less well defined ---and finite--- equilibrium bond lengths. Even for such a large `vessel' as the Earth atmosphere, diatomic N$_2$ or O$_2$ molecules do not conform with the predictions made above. So, what is happening? Which is the ingredient that is missing from the previous analysis in order for it to produce results that are applicable not to interstellar space but to gases here at home?

The answer is simple and clear. We have assumed an infinite volume ---ultimately expressed by the integration in eq.~(\ref{eq:r_av_sph}) from $0$ to $\infty$--- and this is not, in general, valid. Intuitively, if the volume is taken to be finite, since the regions of low $V(\rho)$ contribute with an exponentially larger value under the integral signs in eq.~(\ref{eq:r_av_sph}) than the ones in which $V(\rho)$ is close to its asymptote, there will exist a very large range of total volumes (hand-wavingly, a range that is exponentially large on the depth of the potential well) in which the value of both integrals will change insignificantly little. Of course, for some volume size, the exponentially small areas under the curve will begin to add up and the integrals will begin to depend on the volume. This could be important if it happens for experimentally relevant volumes, but it is entirely irrelevant for typical laboratory setups if the volume for which $\langle \rho \rangle$ starts to grow again is, say, of the order of the diameter of the Milky Way.

We will see in what follows that Nature provides us with examples of both behaviours.

\section{First approximation: Square potential and cutoff on the bond-length}
\label{sec:1st_approximation}

In order to perform a first approximation to the finite-volume case, let us introduce the very simple potential
\begin{equation}
\label{eq:Vsquare}
V(\rho) :=
\begin{cases}
-D & \quad \mathrm{if} \quad \rho \in [0,R] \\
0 & \quad \mathrm{if} \quad \rho > R \\
\end{cases} \ .
\end{equation}

If we introduce it in eq.~(\ref{eq:r_av_sph}), we can analytically compute all integrals if we substitute the infinite upper limit on $\rho$ by a cutoff length $L > R$ (which plays the role of the finite volume available to the molecule in this simple approximation):
\begin{eqnarray}
\label{eq:expr_simplest_Vsquare}
\langle \rho \rangle & = & 
\frac{\int_0^R e^{\beta D} \, \rho^3d\rho + \int_R^L \rho^3d\rho}
     {\int_0^R e^{\beta D} \, \rho^2d\rho + \int_R^L \rho^2d\rho} =
\frac{e^{\beta D} \frac{R^4}{4} + \frac{L^4}{4} - \frac{R^4}{4}}
     {e^{\beta D} \frac{R^3}{3} + \frac{L^3}{3} - \frac{R^3}{3}} \nonumber \\
 & = & \frac{3}{4} \cdot \frac{R^4 (e^{\beta D} - 1) + L^4}
	                          {R^3 (e^{\beta D} - 1) + L^3} =
	   \frac{3}{4} \cdot \frac{R^4 + L^4 / (e^{\beta D} - 1)}
  	 	                      {R^3 + L^3 / (e^{\beta D} - 1)} \ ,
\end{eqnarray}
and we can see that, as we advanced, for not very large $L$ and deep enough wells (as compared to $k_\mathrm{B}T=1/\beta$), the division by $e^{\beta D} - 1$ will kill the dependence on the cutoff $L$, and we will have $\langle \rho \rangle \simeq 3R/4$.

Let us now advance a little bit more in the quantitative characterization and write:
\begin{eqnarray}
\label{eq:expr_simplest_Vsquare_Taylor_1}
\langle \rho \rangle & = & 
  \frac{3}{4} R + \frac{3}{4} \cdot
  \left( \frac{R^4 + L^4 / (e^{\beta D} - 1)}
   	 	      {R^3 + L^3 / (e^{\beta D} - 1)} - R \right) \nonumber \\
 & = & 
 \frac{3}{4} R + \frac{3}{4} \cdot
 \left( \frac{L^3 (L - R) / (e^{\beta D} - 1)}
  	 	     {R^3 + L^3 / (e^{\beta D} - 1)} \right) =:
 \frac{3}{4} R + \frac{3}{4} (L - R) \cdot
         \frac{\epsilon}{1 + \epsilon} \ ,
\end{eqnarray}
where, of course, we have defined in the last step
\begin{equation}
\label{eq:Vsquare_epsilon}
\epsilon := \frac{L^3}{R^3} \frac{1}{e^{\beta D} - 1} \ .
\end{equation}

This is the essential quantity that controls the behaviour of $\langle \rho \rangle$. Indeed, it is easy to see from eq.~(\ref{eq:expr_simplest_Vsquare_Taylor_1})
\begin{enumerate}
\item that $\langle \rho \rangle \to (3/4)R$ if $\epsilon \to 0$, and the aforementioned behaviour with $e^{\beta D} - 1$ and $L$ is now more rigorously represented and controlled by the precise combination in which these two quantities occur in $\epsilon$;
\item and that $\langle \rho \rangle \to (3/4)L$ if $\epsilon \to \infty$, recovering the divergent behaviour when $L^3/R^3$ is much larger than $e^{\beta D} - 1$, i.e., when the volume available to the molecule tends to infinity.
\end{enumerate}

Let us now investigate the regime in which $\epsilon$ is small and Taylor-expand:
\begin{equation}
\label{eq:expr_simplest_Vsquare_Taylor_2}
\frac{\epsilon}{1 + \epsilon} = 0 
 + \left[ \frac{1}{(1 + \epsilon^\prime)^2} \right]_{\epsilon^\prime=0} 
          \epsilon
 - \left[ \frac{1}{(1 + \epsilon^\prime)^3} \right]_{\epsilon^\prime=0} 
   \epsilon^2 + O(\epsilon^3) =
 \epsilon - \epsilon^2 + O(\epsilon^3) \ .
\end{equation}

If we now take the first order of this to eq.~(\ref{eq:expr_simplest_Vsquare_Taylor_1}),
\begin{equation}
\label{eq:eq:expr_simplest_Vsquare_Taylor_3}
\langle \rho \rangle \simeq \frac{3}{4} R
 \left( 1 + \frac{L^3(L-R)}{R^4} \frac{1}{e^{\beta D} - 1} \right) \ ,
\end{equation}
we can see that the dependence of the expected value of the bond length on the cutoff $L$ (the linear volume substitute in this first approximation) will be negligible as long as
\begin{equation}
\label{eq:eq:expr_simplest_Vsquare_condition_1}
\frac{L^3(L-R)}{R^4} \frac{1}{e^{\beta D} - 1} \ll 1 \ ,
\end{equation}
i.e., as long as
\begin{equation}
\label{eq:eq:expr_simplest_Vsquare_condition_2}
\frac{L^4 + R^4 - RL^3}{R^4} \ll e^{\beta D} \ ,
\end{equation}
which, given the fact that we can normally assume $L \gg R$, reduces to
\begin{equation}
\label{eq:eq:expr_simplest_Vsquare_condition_3}
\frac{L^4}{R^4} \ll e^{\beta D}
\quad \Longleftrightarrow \quad
L \ll R e^{\beta D / 4} \ .
\end{equation}

We can now quantitatively see what we advanced before, i.e., that no dependence on the cutoff (on the volume) will be noticed until we go to a value of $L$ that grows exponentially large with the depth of the potential well. More specifically, for each $4 k_B T$ the deeper the well is, we can go to a volume which is $e$ times larger; or for each $8 k_B T$ the deeper the well is, we can go to a volume which is $e^2 \simeq 10$ times larger.

In a typical chemical case, in which we could assume that $R$ is of the order of 1 nm, it would take a well $96 k_B T$ deep (approximately 50 kcal/mol at room temperature) to be able to state that, as long as $L$ is less than 1 km, we will not see any volume-dependence in the equilibrium average of the bond-length $\rho$. For comparison, consider that the well depth for the H$_2$ molecule is approximately $175 k_B T$ at room temperature, which yields a relative independence of the volume for $L$ as large as the diameter of the Solar System (see the next sections for a more in depth discussion of this issue).

\section{Exact finite-volume calculation}
\label{sec:finite_volume}

The cutoff $L$ on the integration range over $\rho$ in eq.~(\ref{eq:expr_simplest_Vsquare}) is just an approximate way of taking into account the finite-volume condition. Obviously, if we want to perform an exact calculation, the existence of a finite box couples the integration range of all coordinates in the relative frame of reference, as we will see in detail what follows.

The starting point for the exact calculation is eq.~(\ref{eq:r_av_Euc}), where the integration is still expressed in terms of the atomic Euclidean coordinates $\vec{q}_a:=(x_a^1,x_a^2,x_a^3)$ and $\vec{q}_b:=(x_b^1,x_b^2,x_b^3)$. 

It is tempting now to change coordinates to the center of mass and the relative position $\vec{r} := \vec{q}_b - \vec{q}_a$ (see, e.g., chap.~VII of \cite{Cohen-Tannoudji1977}). This is very useful if we are interested in the dynamics of the system, but, in the context of equilibrium statistical mechanics, we can get away with performing a simpler change (similar to what is done in \cite{Echenique2006}) in which the new coordinates that describe the overall translation of the molecule are not the center-of-mass ones, but just the coordinates of one of the particles, say, $\vec{q}_a$ and the relative position $\vec{r}$. This would produce a not so nice expression for the kinetic energy as compared with the center-of-mass case, but we can forget about this problem because the momenta can be eliminated from the statistical mechanics objects \emph{before} performing the change (as long as we are interested only on the expected value of momenta-independent observables).

More explicitly, the change of coordinates is as follows:
\begin{subequations}
\label{eq:change_of_coordinates_1mol}
\begin{align}
\vec{R} & = \vec{q}_a \ , \label{eq:change_of_coordinates_1mol_a} \\
\vec{r} & = \vec{q}_b - \vec{q}_1 \ , \label{eq:change_of_coordinates_1mol_b}
\end{align}
\end{subequations}
with inverse
\begin{subequations}
\label{eq:change_of_coordinates_inv_1mol}
\begin{align}
\vec{q}_a & = \vec{R} \ , \label{eq:change_of_coordinates_inv_1mol_a} \\
\vec{q}_b & = \vec{R} + \vec{r} \ , 
\label{eq:change_of_coordinates_inv_1mol_b}
\end{align}
\end{subequations}
and whose Jacobian matrix is
\begin{equation}
\label{eq:J_1mol}
J := \left(\frac{\partial (\vec{q}_a,\vec{q}_b)}
                {\partial (\vec{R},\vec{r})}\right) =
\left( \begin{array}{ccc|ccc}
1 & 0 & 0 & 0 & 0 & 0 \\
0 & 1 & 0 & 0 & 0 & 0 \\
0 & 0 & 1 & 0 & 0 & 0 \\
\hline
1 & 0 & 0 & 1 & 0 & 0 \\
0 & 1 & 0 & 0 & 1 & 0 \\
0 & 0 & 1 & 0 & 0 & 1
\end{array} \right) \ ,
\end{equation}
with
\begin{equation}
\label{eq:detJ_1mol}
\det J = 1 \ .
\end{equation}

This is good news for the change of coordinates under the integral sign in eq.~(\ref{eq:r_av_Euc}), but there is still some complications associated to the changes in the integration limits. 

The first way that comes to mind to change these limits is to argue that, first, since $\vec{q}_a = \vec{R}$, each one of the components of $\vec{R}$, which we will denote as $\vec{R} = (X,Y,Z)$, can in principle vary from $0$ to $L$, \emph{but}, for each value of, say, $X$, the corresponding component in $\vec{r} = (x,y,z)$ must range in a given way if we want that, by virtue of eq.~(\ref{eq:change_of_coordinates_inv_1mol_b}), $x_b^1 = X + x$ (the $x$ component of $\vec{q}_b$) stays in $[0,L]$. Specifically, we must require that $x$ ranges from $-X$ to $L-X$, and the integrals in eq.~(\ref{eq:r_av_Euc}) become
\begin{equation}
\label{eq:int_fv_1}
\int_0^L \int_0^L \int_0^L dXdYdZ
  \int_{-X}^{L-X} \int_{-Y}^{L-Y} \int_{-Z}^{L-Z}
  \rho^k e^{-\beta V(\rho)} \, dxdydz \ ,
\end{equation}
where $k=1$ for the numerator in eq.~(\ref{eq:r_av_Euc}) and $k=0$ for the denominator.

The problem with this expression is that, since the integration limits for $\vec{r}$ depend on $\vec{R}$ and not the other way around, we have to perform the integral on $\vec{r}$ first and we cannot eliminate the $\vec{R}$ coordinates easily. In order to find the alternative expression that allows to do this, we can take a look at eq.~(\ref{eq:change_of_coordinates_1mol_b}) to see that, in principle, say, the component $x$ of $\vec{r}$ can range from $-L$ to $L$, \emph{but then}, for each value of $x$, the corresponding component $X$ in $\vec{R}$ cannot range from $0$ to $L$, or we will have that, according to eq.~(\ref{eq:change_of_coordinates_inv_1mol_b}), in this case, $x_b^1 = X + x$ would move out of $[0,L]$. To avoid this, it may seem that we should ask $X$ to range in $[-x,L-x]$, but $X=x_a^1$ and hence it cannot be negative, nor larger than $L$. Therefore, for $x<0$, we have the range $[-x,L]$, and, for $x>0$, we have $[0,L-x]$\footnote{\label{foot:convince} The reader can convince herself that these ranges are correct by drawing the 2D-graph of the integration region in the orthogonal axes $(x_a^1,x_b^1)$ and then performing the change. The author of this manuscript actually needed to do this.}. This forces us to divide the integrals as follows:
\begin{eqnarray}
\label{eq:integrals_division}
\int_{-L}^L dx \int dX & = &
\int_{-L}^0 dx \int_{-x}^L dX + \int_0^L dx \int_0^{L-x} dX \nonumber \\
 & = & \int_{-L}^0 (L+x) \, dx + \int_0^L (L-x) \, dx =
       \int_{-L}^L (L-|x|) \, dx \ ,
\end{eqnarray}
where the last line of equalities is only valid if we are integrating objects
that do not depend on $X$. Since $\rho=\sqrt{\vec{r}\cdot\vec{r}}$, this is precisely the case of eq.~(\ref{eq:r_av_Euc}), whose integrals now become:
\begin{equation}
\label{eq:int_fv_2}
I[O(\rho)] := \int_{-L}^L \int_{-L}^L \int_{-L}^L (L-|x|) (L-|y|) (L-|z|)
  \, O(\rho) e^{-\beta V(\rho)} \, dxdydz \ ,
\end{equation}
where we have now decided to substitute $\rho^k$ by the more general function $O(\rho)$ on the bond-length.

We could now perform a change to spherical coordinates on the relative position $\vec{r}$:
\begin{subequations}
\label{eq:spherical_inv}
\begin{align}
x & = \rho \sin \theta \cos \phi \ , \label{eq:spherical_inv_x} \\
y & = \rho \sin \theta \sin \phi \ , \label{eq:spherical_inv_y} \\
z & = \rho \cos \theta \ , \label{eq:spherical_inv_z} \\
\end{align}
\end{subequations}
append it to eq.~(\ref{eq:change_of_coordinates_inv_1mol}) and thus finally get back to eq.~(\ref{eq:change_of_coords}).

The problem is that we cannot analytically perform the integration on the angles $\theta$ and $\phi$ at eq.~(\ref{eq:int_fv_2}) (after the change of coordinates and the introduction of the corresponding Jacobian determinant $\det J = \rho^2 \sin \theta$) because of two main difficulties:
\begin{itemize}
\item The integration region is not a sphere anymore; it is now the cube $C := [-L,L] \times [-L,L] \times [-L,L]$. This makes the integration limits on the angles $\theta$ and $\phi$ entangled and unmanageable. One thing we could do is to sandwich the cube $C$ between a sphere $S_L$ of radius $L$, which is entirely contained in the cube, and a sphere $S_{\sqrt{3}L}$ of radius $\sqrt{3}L$, which contains $C$. If $O(\rho)$ is positive everywhere, we will have $\int_{S_L} \leq \int_C \leq \int_{S_{\sqrt{3}L}}$, maybe allowing us to produce some useful bounds.
\item Even if we arrive to something useful using the idea in the previous point, we still have to deal with the factors $(L-|x|) (L-|y|) (L-|z|)$, which do not seem to be willing to disappear without some pain if we use spherical coordinates.
\end{itemize}

Henceforth, for general $V(\rho)$ and $O(\rho)$, and without further assumptions, numerical integration seems the only way ahead. But, since that is always a possibility, let us get now a little bit more particular to analytically arrive to some useful approximations which are somehow more advanced than the one in sec.~\ref{sec:1st_approximation}.

\section{Square potential again}
\label{sec:square_again}

\subsection{Without a repulsive core}
\label{subsec:without_core}

Encouraged by the success of the simple solution in sec.~\ref{sec:1st_approximation}, let us try again the square potential in eq.~(\ref{eq:Vsquare}) but this time using the \emph{correct} integral in eq.~(\ref{eq:int_fv_2}).

If we denote by $S_R$ the sphere with radius $R$ and centered in $(0,0,0)$, and by $W := C \smallsetminus S_R$ the rest of the volume of the cube (we will assume that $S_R$ is entirely contained in $C$), we can insert the square potential into eq.~(\ref{eq:int_fv_2}) to yield
\begin{eqnarray}
\label{eq:int_fv_square_1}
I[O(\rho)] & = &
   e^{\beta D} \int_{S_R} (L-|x|) (L-|y|) (L-|z|) \, O(\rho) \, dxdydz \nonumber \\
 & & \mbox{} + \int_W (L-|x|) (L-|y|) (L-|z|) \, O(\rho) \, dxdydz \ .
\end{eqnarray}

In order to proceed further, we need to perform some reasonable, additional approximations. First, if $L \gg R$ (i.e., the vessel is not of molecular size), we have that, in the first integral, $(L-|x|) (L-|y|) (L-|z|) \simeq L^3$ in the whole integration region $S_R$. Second, and under the same condition that $L \gg R$, we have the volume of $S_R$ is much smaller than the volume of the whole cube $C$. Since the integrand of the second integral is positive, and not very different inside $S_R$ than outside it, this relation between the volumes lets us approximately extend the integral to the whole space $C$. In fact, this amounts to adding
\begin{equation}
\label{eq:int_fv_square_1_error}
\int_{S_R} (L-|x|) (L-|y|) (L-|z|) \, O(\rho) \, dxdydz \ ,
\end{equation}
which is much smaller than the first term above because of the factor $e^{\beta D}$ (i.e., we add the assumption that $D \gg k_\mathrm{B}T$ to the mix), and much smaller than the second one because of the assumption that $L \gg R$. Of course, we have also assumed that $O(\rho)$ is sufficiently well behaved.

Putting all together, we have
\begin{eqnarray}
\label{eq:int_fv_square_2}
I[O(\rho)] & \simeq & e^{\beta D} L^3 \int_{S_R} O(\rho) 
                      \, \rho^2 \sin \theta d\rho d\theta d\phi \nonumber \\
 & & \mbox{} + \int_C (L-|x|) (L-|y|) (L-|z|) \, O(\rho) \, dxdydz \nonumber \\
 & = & 4\pi e^{\beta D} L^3 \int_0^R O(\rho) \, \rho^2 d\rho
       + \int_C (L-|x|) (L-|y|) (L-|z|) \, O(\rho) \, dxdydz \ .
\end{eqnarray}

And we can now take this result to the general expression for $\langle \rho \rangle$ in eq.~(\ref{eq:r_av_Euc}), yielding:
\begin{equation}
\label{eq:r_av_square_1}
\langle r \rangle \simeq
 \frac{4\pi e^{\beta D} L^3 \int_0^R \rho^3 \, d\rho +
       \int_C (L-|x|) (L-|y|) (L-|z|) \sqrt{x^2+y^2+z^2} \, dxdydz}
      {4\pi e^{\beta D} L^3 \int_0^R \rho^2 \, d\rho +
	   \int_C (L-|x|) (L-|y|) (L-|z|) \, dxdydz} \ .
\end{equation}

The integrals over $S_R$ are both of them trivial. The integral over $C$ in the denominator is simple too:
\begin{eqnarray}
\label{eq:intC_1}
\int_C (L-|x|) (L-|y|) (L-|z|) \, dxdydz & = & 
 \int_{-L}^L (L-|x|) \, dx \int_{-L}^L (L-|y|) \, dy \int_{-L}^L (L-|z|) \, dz
  \nonumber \\
 & = & 2^3 \int_0^L (L-x) \, dx  \int_0^L (L-y) \, dy  \int_0^L (L-z) \, dz
 \nonumber \\
 & = & 2^3 \left( \frac{L^2}{2} \right)^3 = L^6 \ ,
\end{eqnarray}
where we have used that the integrands on the right-hand side of the first line are even functions of $x$, $y$ and $z$ in the integration region.

However, the second integral in the numerator is complicated, so we can use that $\langle \rho \rangle \simeq \sqrt{\langle \rho^2 \rangle}$ for small fluctuations, and calculate $\langle \rho^2 \rangle$ instead. Then, the second integral in the denominator becomes:
\begin{eqnarray}
\label{eq:intC_2}
 &   & \int_C (L-|x|) (L-|y|) (L-|z|) (x^2+y^2+z^2) \, dxdydz \nonumber \\
 & = & 8 \int_0^L \int_0^L \int_0^L (L-x) (L-y) (L-z) (x^2+y^2+z^2) \, dxdydz \nonumber \\
 & = & 8 \left( \int_0^L x^2 (L-x) dx \int_0^L (L-y) dy \int_0^L (L-z) dz \right)
       \nonumber \\
 & & \mbox{} + 8 \left( \int_0^L (L-x) dx \int_0^L y^2 (L-y) dy \int_0^L (L-z) dz \right)
       \nonumber \\
 & & \mbox{} + 8 \left( \int_0^L (L-x) dx \int_0^L (L-y) dy \int_0^L z^2 (L-z) dz \right)
       \nonumber \\
 & = & 24 \left( \int_0^L x^2 (L-x) dx \int_0^L (L-y) dy \int_0^L (L-z) dz \right)
       \nonumber \\
 & = & 24 \left( \frac{L^4}{3} - \frac{L^4}{4} \right)
         \left( \frac{L^2}{2} \right) \left( \frac{L^2}{2} \right)
		 = \frac{L^8}{2} \ ,
\end{eqnarray}
and we can take this result and the previous one to $\langle \rho^2 \rangle$:
\begin{eqnarray}
\label{eq:r2_av_square_1}
\langle \rho^2 \rangle & \simeq &
 \frac{4\pi e^{\beta D} L^3 \int_0^R \rho^4 \, d\rho +
       \int_C (L-|x|) (L-|y|) (L-|z|) (x^2+y^2+z^2) \, dxdydz}
      {4\pi e^{\beta D} L^3 \int_0^R \rho^2 \, d\rho +
	   \int_C (L-|x|) (L-|y|) (L-|z|) \, dxdydz} \nonumber \\
 & = & \frac{\displaystyle \frac{4\pi}{5} R^5 L^3 e^{\beta D} + \frac{L^8}{2}}
      {\displaystyle \frac{4\pi}{3} R^3 L^3 e^{\beta D} + L^6} = 
 \frac{\displaystyle \frac{3}{5} R^5 e^{\beta D} + \frac{3}{8\pi}L^5}
	      {\displaystyle R^3 e^{\beta D} + \frac{3}{4\pi} L^3} \ .
\end{eqnarray}

The first thing we notice from this expression is that, again, the bond length diverges if we go to an infinitely large box; i.e.,
\begin{equation}
\label{eq:r2_av_square_1_limL}
\lim_{L \to \infty} \langle \rho^2 \rangle
 = \lim_{L \to \infty} \frac{L^2}{2}
 = \infty \ .
\end{equation}

Second, we see that,
\begin{equation}
\label{eq:r2_av_square_1_limbetaD}
\lim_{\beta D \to \infty} \langle \rho^2 \rangle
 = \frac{3}{5} R^2 \ ,
\end{equation}
which suggests that we should rewrite eq.~(\ref{eq:r2_av_square_1}) as
\begin{eqnarray}
\label{eq:r2_av_square_1bis}
\langle \rho^2 \rangle & \simeq & \frac{3}{5} R^2 +
 \frac{\displaystyle \frac{3}{5} R^5 e^{\beta D} + \frac{3}{8\pi}L^5
       - \frac{3}{5} R^5 e^{\beta D} - \frac{9}{20\pi} R^2 L^3}
	      {\displaystyle R^3 e^{\beta D} + \frac{3}{4\pi} L^3} \nonumber \\
 & = & \frac{3}{5} R^2 +
 \frac{\displaystyle \frac{3}{8\pi}L^5 - \frac{9}{20\pi} R^2 L^3}
      {\displaystyle R^3 e^{\beta D} + \frac{3}{4\pi} L^3} =
 \frac{3}{5} R^2 + \left( \frac{1}{2} L^2 - \frac{3}{5} R^2 \right)
  \frac{\epsilon}{1 + \epsilon} \ ,
\end{eqnarray}
where we have defined
\begin{equation}
\label{eq:epsilon_square_1}
\epsilon = \frac{3}{4\pi} \frac{L^3}{R^3} e^{- \beta D} \ .
\end{equation}

As it happened with the analogous and very similar quantity in eq.~(\ref{eq:Vsquare_epsilon}) in sec.~\ref{sec:1st_approximation}, $\epsilon$ now controls the different regimes for the behaviour of $\langle \rho^2 \rangle$. Indeed, we can see again that
\begin{enumerate}
\item $\langle \rho^2 \rangle \to (3/5)R^2$ if $\epsilon \to 0$; i.e., $\langle \rho \rangle \simeq \sqrt{\langle \rho^2 \rangle} \to \sqrt{3/5}R$ if $\epsilon \to 0$;
\item and $\langle r^2 \rangle \to L^2/2$ if $\epsilon \to \infty$, reproducing the divergent behaviour when $L^3/R^3$ is much larger than $e^{\beta D}$ that we saw in sec.~\ref{sec:1st_approximation}.
\end{enumerate}

We can now investigate the behaviour for small $\epsilon$ and use the exact same Taylor expansion in eq.~(\ref{eq:expr_simplest_Vsquare_Taylor_2}) to substitute for the first order and arrive to
\begin{equation}
\label{eq:r2_av_square_1ter}
\langle \rho^2 \rangle \simeq \frac{3}{5} R^2 
 \left( 1 + \left[\frac{5}{6} \frac{L^2}{R^2} - 1\right]
  \frac{3}{4\pi} \frac{L^3}{R^3} e^{- \beta D} \right) \simeq
 \frac{3}{5} R^2 \left( 1 + \frac{5}{8\pi} \frac{L^5}{R^5}
   e^{- \beta D} \right) \ ,
\end{equation}
where, in the last step, we have used again the fact that we are working in the $L \gg R$ (box much larger than molecule) regime.

Finally, we see that, disregarding numerical factors of the order of unity, we will have an approximately $L$-independent expected value $\langle \rho^2 \rangle = (3/5)R^2$ as long as
\begin{equation}
\label{eq:square_condition_1}
\frac{L^5}{R^5} e^{- \beta D} \ll 1 \quad \Longleftrightarrow \quad
L^5 \ll R^5 e^{\beta D}             \quad \Longleftrightarrow \quad 
L \ll R e^{\frac{\beta D}{5}} \ ,
\end{equation}
which is again very similar to what we found in sec.~\ref{sec:1st_approximation}.

We also remind the reader that we have assumed two additional conditions which are related to the previous one, but which are in part independent and needed for the validity of the analysis in this section (in particular, to be able to write eq.~(\ref{eq:int_fv_square_2})), namely, that
\begin{subequations}
\label{eq:square_condition_1bis}
\begin{align}
L & \gg R \ ,  \label{eq:square_condition_1bis_a} \\
e^{- \beta D} & \ll 1 \ .  \label{eq:square_condition_1bis_b}
\end{align}
\end{subequations}

\subsection{With a repulsive core}
\label{subsec:with_core}

Due to the fact that nuclei are positively charged particles, when they get very close the Coulombic repulsion force between them becomes divergent. This is why it is customary to consider model potentials such as the Morse one in sec.~\ref{sec:H} which present a repulsive core. In this section, we will repeat the previous calculations with a slightly more general potential than the one in eq.~(\ref{eq:Vsquare}) including this feature:
\begin{equation}
\label{eq:Vsquare_2}
V(\rho) :=
\begin{cases}
\infty & \quad \mathrm{if} \quad \rho \in [0,R_1] \\
-D & \quad \mathrm{if} \quad \rho \in (R_1,R_2] \\
0 & \quad \mathrm{if} \quad \rho \in (R_2,\infty) \\
\end{cases} \ .
\end{equation}

Now, we will consider the basic kind of integral in eq.~(\ref{eq:int_fv_2}) and rewrite it using a slightly more compact notation:
\begin{equation}
\label{eq:int_fv_3}
I[O(\rho)] = \int_C \alpha(\vec{r}) \, O(\rho) \, e^{-\beta V(\rho)} \, dv \ ,
\end{equation}
where we have defined
\begin{equation}
\label{eq:alpha}
\alpha(\vec{r}) := (L - |x|) (L - |y|) (L - |z|) \ ,
\end{equation}
the integration region is again given by the cube $C := [-L,L] \times [-L,L] \times [-L,L]$, the Euclidean volume element is $dv:=dxdydz$, and $O(\rho)$ is in principle a general function only of the bond-length $\rho$.

If we now denote by $S_1$ the sphere with radius $R_1$ and centered in $(0,0,0)$, by $S_{12}$ the spherical shell of all the points with $r \in (R_1,R_2]$, by $S_2 := S_1 \cup S_{12}$ its union, and by $W := C \smallsetminus S_2$ the rest of the volume of the cube $C$ (we will assume that $S_2$ is entirely contained in $C$), we can easily divide the integral in eq.~(\ref{eq:int_fv_3}) for the case of the potential in eq.~(\ref{eq:Vsquare_2}):
\begin{equation}
\label{eq:int_fv_square_3}
I[O(\rho)] = e^{\beta D} \int_{S_{12}} \alpha(\vec{r}) \, O(\rho) \, dv
         + \int_W \alpha(\vec{r}) \, O(\rho) \, dv \ .
\end{equation}

Similarly to what we did in the previous sections, the first approximation that we will perform here is to assume that $R_2 \ll L$ (the vessel is much larger than the molecule), which allows us to substitute $\alpha(\vec{r}) \simeq L^3$ in the first integral above:
\begin{equation}
\label{eq:int_fv_square_4}
I[O(\rho)] \simeq L^3 e^{\beta D} \int_{S_{12}} O(\rho) \, dv
         + \int_W \alpha(\vec{r}) \, O(\rho) \, dv \ .
\end{equation}

As we did before, we can use again that $R_2 \ll L$ to argue that $S_2$ is a much smaller volume than $C$, in such a way that, for integrands which do not depend very strongly on $\rho$ (especially in $S_2$), we can substitute integrals over $W$ by integrals over $C = W \cup S_2$. Again, this amounts to adding
\begin{equation}
\label{eq:int_fv_square_4_error}
\int_{S_2} \alpha(\vec{r}) \, O(\rho) \, dv \ ,
\end{equation}
which, for well behaved functions $O(\rho)$, is much smaller than the first term above if we assume that $e^{-\beta D} \ll 1$, and much smaller than the second one because of the already made assumption that $L \gg R_2$.

In order to complete the calculation of the average value $\langle \rho^2 \rangle$, we now follow the same steps as in the previous section:
\begin{eqnarray}
\label{eq:r2_av_square_2}
\langle \rho^2 \rangle & \simeq &
 \frac{4\pi e^{\beta D} L^3 \int_{R_1}^{R_2} \rho^4 \, d\rho +
       \int_C \alpha(\vec{r}) (x^2+y^2+z^2) \, dv}
      {4\pi e^{\beta D} L^3 \int_{R_1}^{R_2} \rho^2 \, d\rho +
	   \int_C \alpha(\vec{r}) \, dv} \nonumber \\
 & = & \frac{\displaystyle \frac{4\pi}{5} \left( R_2^5-R_1^5 \right) L^3 e^{\beta D}
                           + \frac{L^8}{2}}
      {\displaystyle \frac{4\pi}{3} \left( R_2^3-R_1^3 \right) L^3 e^{\beta D} + L^6} = 
 \frac{\displaystyle \frac{3}{5} \left( R_2^5-R_1^5 \right) e^{\beta D} + \frac{3}{8\pi}L^5}
	      {\displaystyle \left( R_2^3-R_1^3 \right) e^{\beta D} + \frac{3}{4\pi} L^3} \ .
\end{eqnarray}

As it happened for the case without the repulsive core, again, the bond length diverges if we go to an infinitely large box; i.e.,
\begin{equation}
\label{eq:r2_av_square_2_limL}
\lim_{L \to \infty} \langle \rho^2 \rangle
 = \lim_{L \to \infty} \frac{L^2}{2}
 = \infty \ .
\end{equation}

Second, we see that,
\begin{equation}
\label{eq:r2_av_square_2_limbetaD}
\lim_{\beta D \to \infty} \langle \rho^2 \rangle
 = \frac{3}{5} \left( \frac{R_2^5-R_1^5}{R_2^3-R_1^3} \right) =: R_\mathrm{eff}^2 \ ,
\end{equation}
where we have defined
\begin{equation}
\label{eq:Reff}
R_\mathrm{eff} := \sqrt{\frac{3}{5} \left( \frac{R_2^5-R_1^5}{R_2^3-R_1^3} \right)} 
\end{equation}
as the equilibrium bond-length for an infinitely deep well (or zero-$T$).

This suggests again that we should rewrite eq.~(\ref{eq:r2_av_square_2}) as
\begin{eqnarray}
\label{eq:r2_av_square_2bis}
\langle \rho^2 \rangle & \simeq & R_\mathrm{eff}^2 +
 \frac{\displaystyle \frac{3}{5} \left( R_2^5-R_1^5 \right) e^{\beta D} + \frac{3}{8\pi}L^5
       - \frac{3}{5} \left( R_2^5-R_1^5 \right) e^{\beta D}
	   - \frac{3}{4\pi} R_\mathrm{eff}^2 L^3}
   {\displaystyle \left( R_2^3-R_1^3 \right) e^{\beta D} + \frac{3}{4\pi} L^3} \nonumber \\
 & = & R_\mathrm{eff}^2 +
 \frac{\displaystyle \frac{3}{8\pi}L^5 - \frac{3}{4\pi} R_\mathrm{eff}^2 L^3}
      {\displaystyle \left( R_2^3-R_1^3 \right) e^{\beta D} + \frac{3}{4\pi} L^3}
	\nonumber \\
 & = & R_\mathrm{eff}^2 + \left( \frac{1}{2} L^2 - R_\mathrm{eff}^2 \right)
	  \frac{\frac{3}{4\pi} L^3}
      {\displaystyle \left( R_2^3-R_1^3 \right) e^{\beta D} + \frac{3}{4\pi} L^3}
	  \nonumber \\
 & = & R_\mathrm{eff}^2 + \left( \frac{1}{2} L^2 - R_\mathrm{eff}^2 \right)
	  \frac{\epsilon}{1 + \epsilon} \ ,
\end{eqnarray}
where we have defined
\begin{equation}
\label{eq:epsilon_square_2}
\epsilon = \frac{3}{4\pi} \frac{L^3}{\left( R_2^3-R_1^3 \right)} e^{- \beta D} \ .
\end{equation}

The result is now very similar to that in the previous sections:
\begin{enumerate}
\item $\langle \rho^2 \rangle \to R_\mathrm{eff}^2$ if $\epsilon \to 0$; i.e., $\langle \rho \rangle \simeq \sqrt{\langle \rho^2 \rangle} \to R_\mathrm{eff}$ if $\epsilon \to 0$;
\item and $\langle r^2 \rangle \to L^2/2$ if $\epsilon \to \infty$, reproducing the divergent behaviour when $L^3/\left( R_2^3-R_1^3 \right)$ is much larger than $e^{\beta D}$ that we saw previously.
\end{enumerate}

For small $\epsilon$, we can use the Taylor expansion in eq.~(\ref{eq:expr_simplest_Vsquare_Taylor_2}) to substitute for the first order and arrive to
\begin{equation}
\label{eq:r2_av_square_2ter}
\langle \rho^2 \rangle \simeq R_\mathrm{eff}^2 
 \left( 1 + \left[\frac{1}{2} \frac{L^2}{R_\mathrm{eff}^2} - 1\right]
  \frac{3}{4\pi} \frac{L^3}{\left( R_2^3-R_1^3 \right)} e^{- \beta D} \right) \simeq
 R_\mathrm{eff}^2 \left( 1 + \frac{5}{8\pi} \frac{L^5}{\left( R_2^5-R_1^5 \right)}
   e^{- \beta D} \right) \ ,
\end{equation}
where, in the last step, we have used again the fact that we are working in the $L \gg R_2$ (box much larger than molecule) regime.

Finally, we see again that, disregarding numerical factors of the order of unity, we will have an approximately $L$-independent expected value $\langle \rho^2 \rangle = R_\mathrm{eff}^2$ as long as
\begin{equation}
\label{eq:square_condition_2}
\frac{L^5}{\left( R_2^5-R_1^5 \right)} e^{- \beta D} \ll 1 \quad \Longleftrightarrow \quad
L^5 \ll \left( R_2^5-R_1^5 \right) e^{\beta D} \quad \Longleftrightarrow \quad 
L \ll \left( R_2^5-R_1^5 \right)^{1/5} e^{\frac{\beta D}{5}} \ .
\end{equation}

\section{Summary and discussion}
\label{sec:summary}

In this document, we have shown that the prediction of classical statistical mechanics for the equilibrium average of the bond-length $\rho$ for a dilute gas of diatomic molecules is, in general, dependent on the volume $L^3$ available to the gas.

Due to the difficulty of the calculations, we have analytically proved this to be the case for a series of square model potentials and using $\sqrt{\langle \rho^2 \rangle}$ as a suitable approximation to $\langle \rho \rangle$. For different details in the potential energy as a function of $\rho$, the quantitative results are also slightly different. However, the qualitative behavior is the same as long as we stick to model potentials with the right qualitative properties; especially to become constant for large $\rho$. For the sake of simplicity, we will then refer to the particular results established in sec.~\ref{subsec:without_core} for the model potential in eq.~(\ref{eq:Vsquare}):
$$
V(\rho) :=
\begin{cases}
-D & \quad \mathrm{if} \quad \rho \in [0,R] \\
0 & \quad \mathrm{if} \quad \rho > R \\
\end{cases} \ .
$$

If we assume that $R \ll L$ (the very reasonable situation in which the box is much larger than the molecule) and also that $e^{\beta D} \ll 1$ (which is true for most diatomic molecules when the temperature is not extremely low), then the main finding is that, as long as $L \ll R e^{\frac{\beta D}{5}}$, the equilibrium bond-length is approximately constant and equal to
$$
\sqrt{\langle r^2 \rangle} \simeq \sqrt{\frac{3}{5}} R \ . 
$$
As the available volume gets larger and the inequality $L \ll R e^{\frac{\beta D}{5}}$ is no longer satisfied, $\sqrt{\langle r^2 \rangle}$ starts to grow, becoming divergent for an infinite volume.

Since $e^2 \simeq 10$, we will have that, for each 10 $k_B T$ we increase the depth of the potential well additively, the possible region in which the expected value of the bond length is constant as well as the box sizes for which we start to see a volume dependence will increase a factor 10 multiplicatively. 

For example, the well depth for the H$_2$ molecule is approximately $175 k_BT$ at room temperature, and hence $e^{\frac{\beta D}{5}} \simeq 10^{17.5} \simeq 3 \cdot 10^{17}$. If we assume that the well width in this case is of the order of $10^{-9}$ m, we have that the expected value of H$_2$ will remain constant up to containers of the order of $3 \cdot 10^{8}$ m, about the radius of the orbit of the Moon around the Earth.\footnote{\label{foot:4vs5} The reader can check by herself that the difference between this estimation and the one made in sec.~\ref{sec:1st_approximation}, which is the difference between $e^{\frac{\beta D}{5}}$ here and $e^{\frac{\beta D}{4}}$ there, is the consequence of using $\sqrt{\langle r^2 \rangle}$ instead of $\langle r \rangle$. This suggests that this approximation will have to be more carefully studied in further works.}

However, in the case of very weakly bound molecules, such as Mn$_2$ \citep{Tzeli2008}, which has a well depth of approximately $2 k_B T$ at room temperature, we can have $e^{\frac{\beta D}{5}}$ of the order of 1, indicating that the expected value of the bond length will probably depend on the volume of the container for all volume ranges. 

In the intermediate range of well depths, and performing the same kind of hand-waving calculations, we can predict that, assuming a well width of $10^{-9}$, we will need that $e^{\frac{\beta D}{5}} \simeq 10^9$ to start to see volume effects for the convenient human-size containers of the order of 1 m. This corresponds to well depths of approximately $90 k_B T$ or about 50 kcal/mol (200 kJ/mol, 0.1 hartree, 20000 cm$^{-1}$, 2 eV) at room temperature. According to \cite{Luo2007} and assuming that the \emph{bond dissociation energy} is more or less equal to the well depth, we see that diatomic molecules of halogens can start to show finite-bottle effect for container sizes which lie in this human-sized range.

\phantomsection
\addcontentsline{toc}{section}{References}
%\bibliography{/Users/pablo/maletin/temp/workbench/mendeley/library}

\end{document}